\documentclass[aps,prl,showpacs,floatfix,twocolumn,superscriptaddress,longbibliography]{revtex4-2}

\usepackage{todonotes}
\usepackage{graphicx}
\usepackage{subcaption}
\usepackage{dcolumn}
\usepackage{bm}
\usepackage[colorlinks=true, allcolors=blue, hyperfigures=true]{hyperref}

\def\LCCO {$\mathrm{La}_{2-x} \mathrm{Ce}_x \mathrm{CuO}_{4}$} 

\begin{document}
\preprint{APS/123-QED}
\title{Evolution of the magnetic excitations in electron-doped \LCCO}

\author{X. T. Li}
\affiliation{School of Physical Science and Technology, ShanghaiTech University, Shanghai 201210, China}

\author{S. J. Tu}
\affiliation{Beijing National Laboratory for Condensed Matter Physics and Institute of Physics, Chinese Academy of Sciences, Beijing 100190, China}

\author{L. Chaix}

\author{C. Fawaz}
\author{M. d'Astuto}

\affiliation{Univ. Grenoble Alpes, CNRS, Grenoble INP, Institut N\'{e}el, 38000 Grenoble, France}

\author{X. Li}
\affiliation{School of Physical Science and Technology, ShanghaiTech University, Shanghai 201210, China}

\author{F. Yakhou-Harris}
\author{K. Kummer}
\author{N. B. Brookes}
\affiliation{European Synchrotron Radiation Facility (ESRF), B.P. 220, F-38043 Grenoble Cedex, France}

\author{M. Garcia-Fernandez}
\author{K. J. Zhou}
\affiliation{Diamond Light Source, Harwell Campus, Didcot, OX11 0DE, United Kingdom}

\author{Z. F. Lin}
\author{J. Yuan}
\author{K. Jin}
\affiliation{Beijing National Laboratory for Condensed Matter Physics and Institute of Physics, Chinese Academy of Sciences, Beijing 100190, China}
\author{M. P. M. Dean}
\affiliation{Condensed Matter Physics and Materials Science Department, Brookhaven National Laboratory, Upton, New York 11973, USA}
\author{X. Liu}
\email{xliu@shanghaitech.edu.cn}
\affiliation{School of Physical Science and Technology, ShanghaiTech University, Shanghai 201210, China}

\date{\today}

\begin{abstract}
We investigated the high energy spin excitations in electron-doped \LCCO ~(LCCO), a cuprate superconductor, by resonant inelastic x-ray scattering (RIXS) measurements. Efforts were paid to disentangle the paramagnon signal from non-spin-flip spectral weight mixing in the RIXS spectrum at $\bf{Q_{\|}}$ = $(0.6\pi, 0)$ and $(0.9\pi, 0)$ along the (1 0) direction. Our results show that, for doping level \textit{x} from 0.07 to 0.185, the variation of the paramagnon excitation energy is marginal. We discuss the implication of our results in connection with the evolution of the electron correlation strength in this system.

\end{abstract}
\maketitle

The superconducting cuprate families are known for their rich phase diagram. The strong electron-electron correlation, as initially noted by Anderson \cite{Anderson}, is believed to be the key ingredient leading to the high temperature superconducting phase. On the other hand, due to the lack of proper experimental evaluation of the electron correlation strength, it remains a major difficulty in understanding the mechanism of the high temperature superconductivity in these doped cuprates across their phase diagram \cite{Bansil_prb2002,Tremblay_prl2004,Bansil_PRL2006,Bansil_prb2010,Bansil_prb2010_exp}. Resonant inelastic X-ray scattering (RIXS) measurement, a technique which can probe the elementary excitations in a large energy range \cite{Luuk_RevModPhys,Dean_review2015}, is expected to be a promising solution. Particularly, the excitations at higher energies near the zone boundary are dominated by local electron-correlation effects.

Recent progress in RIXS measurements on the cuprate families indeed provide rich information about these correlated electron systems, including magnons \cite{polar_cal_Braicovich_prb2010,Tacon2011,Dean2013,Dean_PRL2013,Tacon2013,Anisotropic2014,polarBi2201,polar2015,polar2018,anisotropic_LSCO2019,Ishii,Lee,hardening2017,LSCO_momentum_polar,Zhang_npj2022}, plasmons \cite{plasmon_Nature,Plasmon_Lin,Kejin_PRL2020}, phonons coupling strongly to the electrons \cite{phonon_PRL2019,phonon_Peng_PRL2020,phonon_Lin2020,phonon_PRR2020,phonon_PRB2022}, etc. In the meantime, the RIXS measurements also bring new puzzles. It is found that the spin excitation persists to high doping level in hole-doped cuprates \cite{Tacon2011,Dean2013,Tacon2013,Anisotropic2014,Meyers_prb2017,LSCO_momentum_polar,lightly_doped,anisotropic_LSCO2019}. Then a surprising observation of significant hardening of the magnetic excitations is reported in the electron-doped cuprates \cite{Ishii,Lee,hardening2017}, marked by about $50\%$ increase ($\sim$150 meV) of the magnetic excitation energy at the zone boundary by doping from 0.04 to 0.147 \cite{Lee}. 

The strength of exchange interaction is considered to be critical for high temperature superconducting $T_{c}$\cite{Anderson2007}. Thus clarifying the nature of magnetism is a key step in understanding unconventional superconductivity in electron doped cuprates. The observation of magnetic hardening upon doping is quite counter-intuitive, and challenges current understanding of the electron-doped cuprates. Naively, a magnetic softening would be expected in a diluted spin system 
\cite{dilute_theory_prb1977,dilute_theory_prb1985,dilute_theory_damping_prb2002}. With the enhanced screening from doped itinerant electrons, the magnetic exchange interaction, born from electron-correlation in the electron-doped cuprates, is generally expected to be weakened. Such expectation seems to be consistent with the falling edge of the superconducting dome upon heavier doping. If the magentic exchange becomes stronger instead, the suppression of superconductivity in the over doped region would be yet harder to comprehend.    

The unusual hardening in the electron-doped cuprates was first explained in terms of the so-called three-site exchange mechanism within a ``local-static" picture \cite{Jia}. Further, it is suggested that the hardening is a natural outcome from a $t-J$ type description \cite{band_renormalise}. Both explanations root in very strong electron correlation on the electron-doped side deep into the phase diagram upon doping. This is quite surprising, and revives the discussion on insulator to strange metal cross-over. For example, Weber \textit{et al.} suggested that the electron-doped cuprates should be considered as weakly correlated Slater insulators \cite{Weber}.

\begin{figure*}[htp]
\includegraphics[width=1\textwidth]{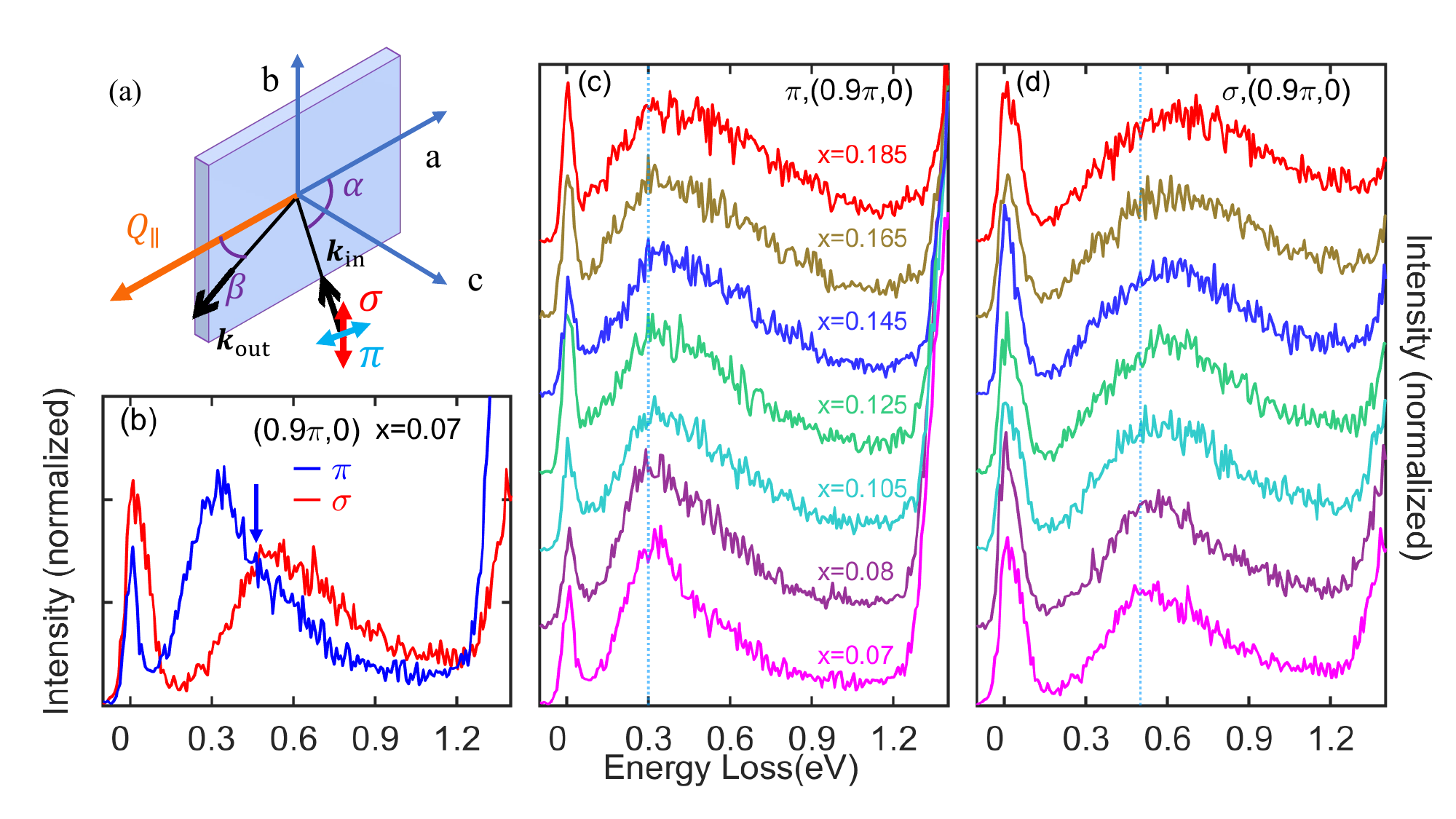}
\caption{\label{fig:geometry_stacking}(a) Experimental geometry. $\alpha$ and $\beta$ are the incident and out-going angles relative to the CuO plane. (b) Example RIXS spectra at $ {\bf Q_{\|}}=(0.9\pi, 0)$, $x=0.07 $. The blue arrow indicates the turning point of the spectrum with $\pi$ incidence. (c)-(d) RIXS spectra taken at ${\bf Q_{\|}}=(0.9\pi,0)$ with incident $\pi$ and $\sigma$ polarization. Dotted lines are guides to the eye.   }
\makeatletter
\let\save@currentlabel\@currentlabel
\edef\@currentlabel{\save@currentlabel(a)}\label{fig:geometry_stacking a}
\edef\@currentlabel{\save@currentlabel(b)}\label{fig:geometry_stacking b}
\edef\@currentlabel{\save@currentlabel(c)}\label{fig:geometry_stacking c}
\edef\@currentlabel{\save@currentlabel(d)}\label{fig:geometry_stacking d}
\makeatother
\end{figure*}

The above puzzle calls for further investigation of the spin excitation in the electron-doped cuprates. It must be noted that the RIXS signal contains excitations of different origins, which could overlap in the spectrum. In particular, it is known that higher order spin excitations always accompany the so-called paramagnon excitations in the RIXS data \cite{Braicovich2010,lightly_doped,multimagnon_prb_2021,multimagnon_PRL2001,multimagnon_prb2012}. Thus the RIXS data interpretation procedures taken in the above mentioned reports \cite{Ishii,Lee,hardening2017} might be over-simplified. For a more proper understanding of the RIXS observation, decoupling of the signals from different origins is required. As we demonstrate in this report, a cross-examination of the RIXS signals from alternation of the incident X-ray polarization for selected ${\bf Q_{\|}}$ points in the reciprocal space can serve this purpose.

In the present work, we investigated the high energy magnetic excitations in the electron doped cuprate \LCCO~(LCCO) as function of doping. Special effort is paid to the RIXS signal decoupling and interpretation.  By cross-examining the signals from both $\sigma$ and $\pi$ incident X-ray channels at particular scattering geometry, the paramagnon signal is well separated out from multimagnon contributions. Results from such analysis show that the spectral weight peak shift in the raw RIXS data is mainly due to the multimagnon contribution. The single spin-flip excitation shows marginal variation upon doping, similar to the hole-doped cuprates \cite{Tacon2011,Dean2013,Tacon2013,Meyers_prb2017,LSCO_momentum_polar,lightly_doped,anisotropic_LSCO2019}.

The RIXS measurements were carried out at beamline ID32 of European Synchrotron Radiation Facility \cite{ESRF_ID32}, and at the I21 beamline at Diamond Light Source, United Kingdom \cite{Diamond_I21}. The incoming beam energy was tuned to the Cu $L_{3}$ edge resonance with an energy resolution of 42 meV (Full Width at Half Maximum). 

Given the highly two-dimensional nature of the spin excitations in the cuprate families \cite{two_dimension}, the momentum transfer perpendicular to the $\text{CuO}_2$ plane is ignored in our measurements. At selected ${\bf Q_{\|}}$ points, namely $(0.6\pi,0)$ and $(0.9\pi,0)$ along the [1 0] direction, RIXS data were taken with the incident X-ray polarization both parallel ($\pi$) and perpendicular ($\sigma$) to the scattering plane (see Fig.~\ref{fig:geometry_stacking a}). The sample used is a high quality doping-concentration-gradient (combi) film \cite{sample,Plasmon_Lin,JinKui_Nature2022}, which allows a doping dependent survey from $\textit{x}=0.07$ to $\textit{x}=0.185$.  All data were recorded at about 18 K.

Our main focus is to utilize the incident X-ray polarization {\it and} scattering geometry dependence of the RIXS scattering cross-section to reliably single out the paramagon signal. These dependencies were discussed by Ament {\it et al.} \cite{single_spin_flip_model_PRL2009} and many others \cite{polar_cal_Braicovich_prb2010,polarBi2201,polar2015,polar2019}. The geometry to discuss is reproduced in  Fig.~\ref{fig:geometry_stacking a}. The most favorable geometry for spin-flip signal is to keep the X-ray exit angle ${\beta}$ low with ${\pi}$ incident polarization \cite{supplement}. Such geometry was employed by many experiments on the cuprates \cite{Tacon2011,Dean2013,Ishii,polar2015,Lee,hardening2017,LSCO_momentum_polar,polar2018}. If the incident X-ray polarization is switched to ${\sigma}$ while keeping the grazing-out setup, the signal from non-spin-flip channel dominates. The discrimination of the spin-flip and non-spin-flip contributions directly depends on the ${\beta}$ angle \cite{supplement}. Thus in our experiments for all data collected, we kept ${\beta}$ angle as low as 5 degrees.

RIXS spectra collected with the above setup are shown in Fig.~\ref{fig:geometry_stacking b}-\ref{fig:geometry_stacking d}. The reliability of our approach is well demonstrated by the data shown in Fig.~\ref{fig:geometry_stacking b}. At ${\bf Q_{\|}}=(0.9\pi,0)$ for {\it x} = 0.07, RIXS signals from $\sigma$ incidence (red) and $\pi$ incidence (blue) are drastically different in the energy range of a few hundreds of meV. As discussed, the $\sigma$ channel is dominated by non-spin-flip multimagnon contribution, and its spectral weight center is at relatively higher energy as expected. For the $\pi$ incidence, a prominent peak appears at $\sim$300 meV, which is assigned to the paramagnon excitation. We emphasize that this peak is riding on significant multimagnon signals. A curvature turning point at $\sim$450 meV (arrow) is visible, which coincides with the peak of the multimagnon signal peak in the $\sigma$ channel. Although the non-spin-flip excitation is suppressed in the $\pi$ channel, it still contributes a significant mixing \cite{polarBi2201,Zhang_npj2022,polar2018}.

Shown in Fig.~\ref{fig:geometry_stacking c}-\ref{fig:geometry_stacking d} is the doping dependent evolution of the RIXS spectra collected in $\pi$ and $\sigma$ channels respectively. By examining only the profile peak position, the spectra shown in Fig.~\ref{fig:geometry_stacking c} in the ${\pi}$ channel (paramagnon signal enhanced) do show a trend of shifting towards higher energy upon heavier doping. While if examining the signal in the $\sigma$ channel where the non-spin-flip contribution dominates, the shifting is even larger. As explained above, the spin-flip signal in the ${\pi}$ channel is riding on significant non-spin-flip spectral weight, the assignment of the peak shifting to paramagon is not straightforward.

 To decouple such mixing and properly single out the paramagon signal, we take the following $\sigma$-$\pi$ channel cross-fitting approach. Both $\sigma$ and $\pi$ channels contain spin-flip and non-spin-flip contributions, which are noted as  $S_{\text{sf}}$ and $S_{\text{nf}}$ respectively. Their mixing ratios are noted as $R_{\text{sf}}$ and $R_{\text{nf}}$. Thus the spin-related excitation signals in both channels in hundreds of meV range can be symbolically written as, 
 
 \begin{equation}
M^{\pi,\sigma} = R_{\mathrm{sf}}^{\pi,\sigma} \cdot S_{\mathrm{sf}} + R^{\pi,\sigma}_{\mathrm{nf}} \cdot S_{\mathrm{nf}} 
\end{equation}
  
Where $M^{\pi,\sigma}$ indicate the spin-related component including paramagnon and multimagnon. As a result, $S_{\mathrm{sf}}$ can be decoupled from,

\begin{equation}
\label{pure magnon}
(R_{\mathrm{sf}}^{\pi} - \frac{R_{\mathrm{nf}}^{\pi}}{R_{\mathrm{nf}}^{\sigma}}R_{\mathrm{sf}}^{\sigma} ) \cdot S_{\mathrm{sf}} = M^{\pi}- \frac{R_{\mathrm{nf}}^{\pi}}{R_{\mathrm{nf}}^{\sigma}} M^{\sigma}
\end{equation}

Although the parameters $R_{\text{sf,nf}}^{\pi,\sigma}$ are unknown, $b = R_{\mathrm{nf}}^{\pi}/R_{\mathrm{nf}}^{\sigma} $ can be treated as a fitting parameter. With $M^{\sigma}$ and $M^{\pi}$ obtained from experimental data, the spectral contribution from paramagnon excitation, namely $S_{\text{sf}}$, can be singled out with a ratio \cite{supplement}. To parameterize the paramagnon spectral contribution $S_{\text{sf}}$, a damped harmonic oscillator model \cite{Dean2013} is used as usual \cite{LSCO_momentum_polar,Meyers_prb2017,polar2018,anisotropic_LSCO2019,theory_dampfunciton2021}, 
$$
f(\omega)=\frac{1}{1-e^{-\beta \hbar \omega}} \frac{4  \omega A_{0} \Gamma_{q}  \sqrt{\omega_q^2+\Gamma_q^2} }{\left(\omega^{2}-\omega_q^{2}-\Gamma_q^2\right)^{2}+(2 \omega \Gamma_{q})^{2}}
$$
 
where the propagating frequency $\omega_{q}$ and damping rate $\Gamma_q$ are the real and imaginary part of the damped oscillator pole \cite{supplement}. Accordingly, the peak maximum $\omega_{\mathrm{max}}$, which is usually larger than $\omega_{q}$ , can be also extracted .

\begin{figure}[htp]
\centering
\includegraphics[width=0.47\textwidth]{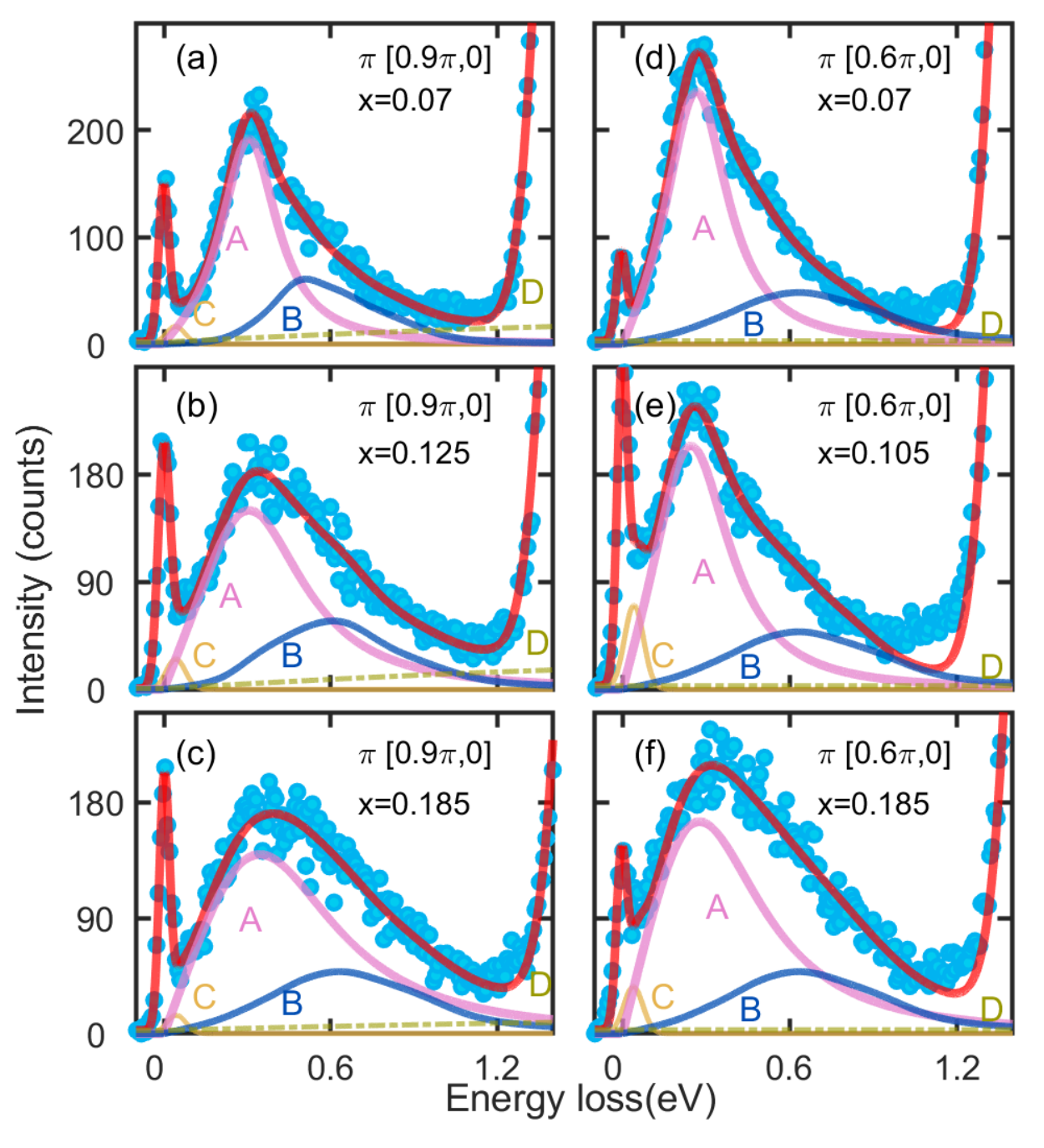}
\caption{\label{fig:fitting} Results from cross-fitting approach. (a)-(c): $\bf{Q}_{\|}$ $= [0.9\pi, 0]$;(d)-(f) $\bf{Q}_{\|}$ $=[0.6\pi, 0] $. The overall spectra are decomposed into elastic peak, paramagnon {\bf A}, multimagnon {\bf B}, phonon {\bf C}, and a small linear background {\bf D}. Dots are the experimental data points.}
\end{figure}

Following the above approach, the fitting results to the experimental data for ${\bf Q_{\|}}$=(0.6$\pi$,0) and (0.9$\pi$,0) at selected doping levels are shown in Fig.~\ref{fig:fitting}. The $\bf B$ component is the line shape of $M^{\sigma}$ scaled by fitting parameter $b$, and the {\bf A} component is the paramagnon as a damped oscillator. Other contributions include elastic peak centered at zero, a small phonon peak {\bf C} and a linear background {\bf D}. More details can be found in the supplementary \cite{supplement}.

\begin{figure*}[htp]
\centering
\includegraphics[width=0.8\textwidth]{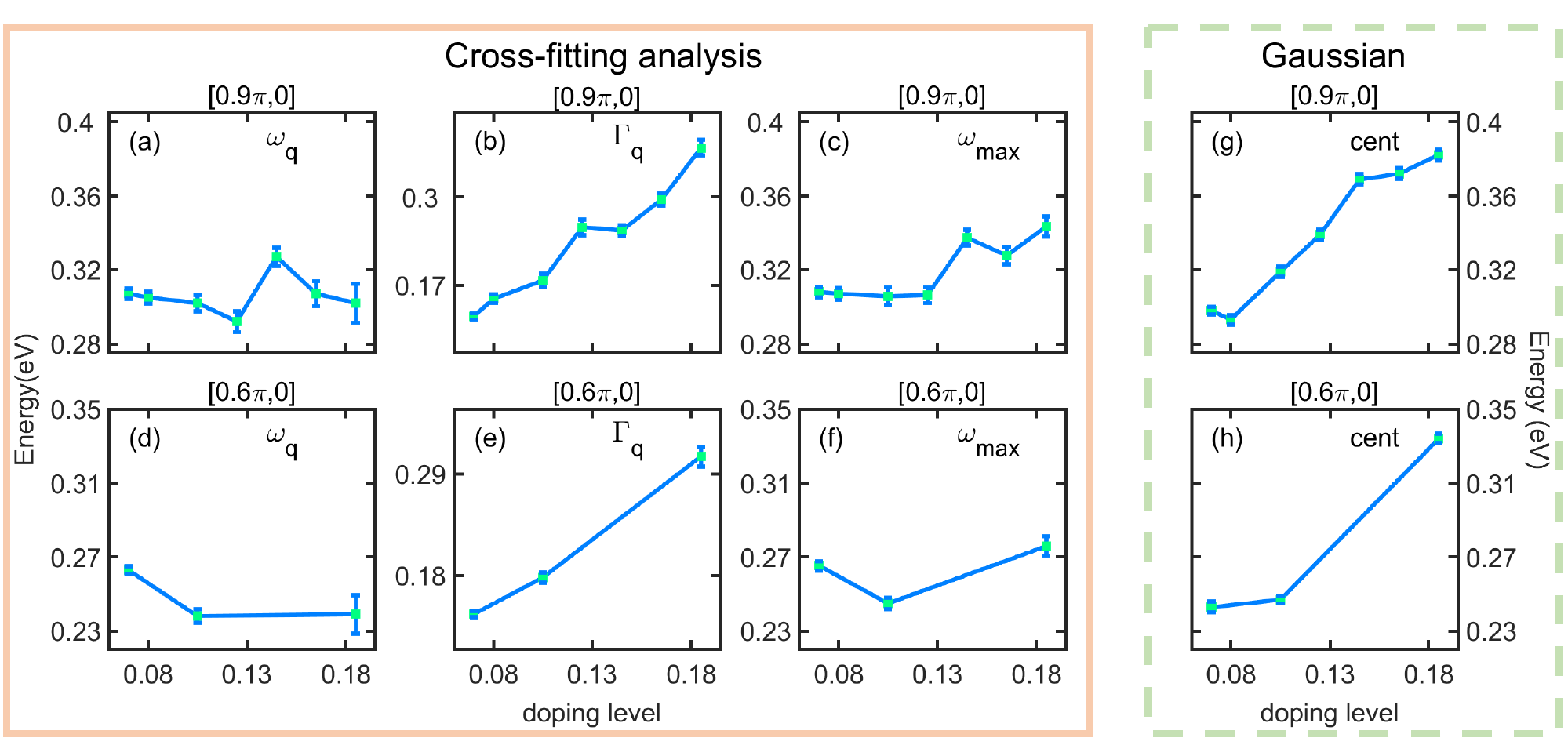}
\caption{\label{fig:magnon_parameter}The evolution of extracted parameters upon doping from the two fitting methods for data at $[0.9\pi,0]$ and $[0.6\pi,0]$. (a)-(f): the propagating frequency $\omega_q$, the damping rate $\Gamma_q$ and the maximum of peak $\omega_{\text{max}}$ from cross-fitting; (g)-(h): the centers of peaks from single Gaussian peak fitting.}
\makeatletter
\let\save@currentlabel\@currentlabel
\edef\@currentlabel{\save@currentlabel(a)}\label{fig:magnon_parameter a}
\edef\@currentlabel{\save@currentlabel(b)}\label{fig:magnon_parameter b}
\edef\@currentlabel{\save@currentlabel(c)}\label{fig:magnon_parameter c}
\edef\@currentlabel{\save@currentlabel(d)}\label{fig:magnon_parameter d}
\edef\@currentlabel{\save@currentlabel(e)}\label{fig:magnon_parameter e}
\edef\@currentlabel{\save@currentlabel(f)}\label{fig:magnon_parameter f}
\edef\@currentlabel{\save@currentlabel(g)}\label{fig:magnon_parameter g}
\edef\@currentlabel{\save@currentlabel(h)}\label{fig:magnon_parameter h}
\makeatother
\end{figure*}

From the cross-fitting results, we can examine the evolution of the high energy paramagnons on a solid basis. Shown in Fig.~\ref{fig:magnon_parameter a}-\ref{fig:magnon_parameter f} are the extracted real part of the propagating frequency $\omega_q$, damping rate $\Gamma_q$ and peak energy $\omega_{\text{max}}$. The most important observation is that $\omega_{q}$ shows marginal variation as function of doping. As to the damping rate $\Gamma_q$, it increases in a linear fashion with doping. This is consistent with the expectation that scattering rate of the magnetic excitations increases upon heavier doping. At low doping, $\Gamma_q$ is significantly smaller than $\omega_{q}$. Thus the paramagnon excitation peak is better defined, as shown in Fig.~\ref{fig:geometry_stacking b}. At high doping {\it x} = 0.185, the damping rate is more than doubled, and the paramagon peak is significantly broadened to merge together with the multimagnon signals.  The increasing of the damping rate also drives the peak maximum towards higher energy moderately, as shown in Fig.~\ref{fig:magnon_parameter c} and \ref{fig:magnon_parameter f}. At the near zone boundary point $[0.9\pi, 0]$, the shift of $\omega_{\text{max}}$ is about 40 meV in our studied doping range.

For the purpose of comparison, we also fitted the data in $\pi$ channel directly with two Gaussian peaks to account for the overall spectral contribution in hundreds of meV range \cite{supplement}. The results are shown in Fig.~\ref{fig:magnon_parameter g}-\ref{fig:magnon_parameter h}. This simplified treatment leads to a ``hardening" of the excitation by $\sim$ 100 meV from {\it x} = 0.07 to 0.185 at both ${\bf Q_{\|}}$ points, similar to earlier reports \cite{Lee,Ishii,hardening2017}. From our above analysis, we now see that about half of this ``hardening" is from multimagnon signal mixing, and the other half is from increased scattering rate of the magnetic excitations upon heavier doping. As to the key parameter, namely the propagating frequency $\omega_{q}$, its variation upon doping is marginal. To further support our results, it is interesting to check earlier neutron studies on the electron doped cuprates \cite{Fujita2011,Fujita2008,NAKAGAWA2008}. There the extracted local spin susceptibility is much weaker than that of the hole-doped counterpart. Although the dispersion cannot be well resolved due to weak inelastic neutron scattering signal, their observed broadening of the peak width upon heavier doping suggests a weak softening of the magnetic excitations. These evidences imply that a strongly localized spin picture might be at failure in describing the spin dynamics in the electron doped cuprates.

It is worth noting that, even with Gaussian peak fitting to the $\pi$-channel only, the extracted paramagnon peak ``hardening" at the zone boundary shown in Fig.~\ref{fig:magnon_parameter g} and \ref{fig:magnon_parameter h} is about {50\%} less than reported earlier \cite{Lee}. This difference clearly emphasizes the significance of the geometry dependent non-spin-flip signal mixing in distorting the RIXS spectrum shape, since our $\pi$ channel data was taken at incident angle ${\beta} = 5^\circ$, different from before. Our cross-channel fitting results lead to a less than 50 meV shifting in paramagnon peak maximum position, and almost non-observable variation in the harmonic model propagating frequency $\omega_q$. This is in contrast to the theoretical analysis invoking the three-site
exchange mechanism \cite{Jia,three_site_prb2019,three_site_prb1997} or doping assisted hopping enhancement in $t-J$ model \cite{band_renormalise,hardening_t_J_prb2005,hardening2020}. We believe that these mechanisms are at play, except that the evolution of the electron-correlation strength in this electron-doped cuprate family is overlooked in the calculations. In this regard, the RIXS measurements on the evolution of spin excitations upon doping in the cuprates should be more informative than previously considered. Once the high energy paramagnon excitation is properly extracted, it should provide strong guidance in evaluating the evolution of electron-correlation for future theoretical studies.

We note that the most rigorous way to decouple spin-flip and non-spin-flip signals is to perform a full X-ray polarization analysis, including incident X-ray polarization control as what we did, {\it{and}} scattered RIXS signal polarization analysis to single out the spin-flip components. This treatment has been demonstrated by \cite{polar2015,polar2018,polar2019}. Since such approach is extremely experimentally demanding, it is difficult to be applied to phase diagram survey.

In summary, we investigate the doping evolution of  spin excitation by Cu $L_{3}$ edge RIXS measurement on the electron-doped cuprate LCCO. Special geometry and polarization conditions are selected to properly separate the paramagnon signal from multimagnon spectral weight mixing. From the decoupled paramagnon signal, we find that the hardening of paramagnon component is insignificant when doping changes from x=0.07 to x=0.185, inconsistent with predictions from three-site
exchange mechanism with a ``local-static” picture \cite{Jia}, or from $t-J$ type picture \cite{band_renormalise}. Such discrepancy is likely due to improper implementation of the evolution of the electron-correlation in this cuprate family.  

We thank Dr. Yifan Jiang and Jian Kang at ShanghaiTech University for valuable discussions. X. T. Li, X. Li and X. Liu acknowledge support from the MOST of China under Grant No. 2022YFA1603900 and from NSFC under Grant No. 11934017. S. J. Tu, Z. F. Lin, J. Yuan and K. Jin acknowledge support from the CAS project for Young Scientists in Basic Research under Grant No. YSBR-048 and from NSFC under Grant No. 11927808. Work at Brookhaven National Laboratory was supported by the U.S. Department of Energy, Office of Science, Office of Basic Energy Sciences. We acknowledge European Synchrotron Radiation Facility for beam time at ID32 under Proposal No. HC-4403, and Diamond Light Source for providing beam time at I21 under proposal MM27478.

\nocite{Achkar_SciRep_IPFY2010,CXRO,Henke_CXRO,FLUO,Sala_2011}
\bibliography{ref}
\end{document}